\documentclass[12pt,showpacs,preprint,eqsecnum,amsmath,amssymb]{revtex4}
\usepackage{graphicx}
\usepackage{amsmath}
\usepackage{amssymb}
\begin{document}
\title{BOUND STATES IN THE CONTINUUM  \\
IN TWO-DIMENSIONAL SERIAL STRUCTURES}
\author{Giorgio Cattapan}
\affiliation{Dipartimento di Fisica ``G. Galilei'', Universit\`a di Padova,
Via F. Marzolo 8, I-35131 Padova, Italy \\
Istituto Nazionale di Fisica Nucleare, Sezione di Padova, 
Via F. Marzolo 8, I-35131 Padova, Italy}
\author{Paolo Lotti}\thanks{I would like to acknowledge the Physics Department
of Padua University for hospitality and support.}
\affiliation{Istituto Nazionale di Fisica Nucleare, Sezione di Padova, 
Via F. Marzolo 8, I-35131 Padova, Italy} 
\begin{abstract}
We investigate the occurrence of bound states in the
continuum (BIC's) in serial structures of quantum dots coupled to an external 
waveguide, when some characteristic length of the system is changed. By 
resorting to a multichannel scattering-matrix approach, we show that BIC's 
do actually occur in two--dimensional serial structures, and that they are a 
robust effect. When a BIC is produced in a two--dot system, it also occurs 
for several coupled dots. We also show that the complex dependence of the 
conductance upon the geometry of the device allows for a simple picture in 
terms of the resonance pole motion in the multi--sheeted Riemann energy 
surface. Finally, we show that in correspondence to a zero--width state for 
the open system one has a multiplet of degenerate eigenenergies for the 
associated closed serial system, thereby generalizing results previously 
obtained for single dots and two-dot devices.
\end{abstract} 

\maketitle
\section{Introduction}
\label{intro}

Since the seminal paper by von Neumann and Wigner~\cite{nw29}, the
occurrence of isolated discrete eigenvalues embedded in the continuum
of scattering states has been the subject of several studies. Bound
states in the continuum (BIC's) have been found by Fonda and Newton
for a model two-channel system with square well potentials \cite{fn60},
and by Friedrich and Wintgen for the hydrogen atom in a uniform
magnetic field \cite{fw85i}. More generally, resorting to Feshbach's
theory of resonances, Friedrich and Wintgen were able to prove that
BIC's can occur because of the interference of resonances belonging
to different channels \cite{fw85ii}. When the relative position of the
resonances changes as a function of a continuous parameter of the system,
their interference produces an avoided crossing of the resonance positions;
at the same time, a dramatic change in their widths occurs, and for a
given value of the parameter one of the resonances acquires an exactly
vanishing width, thereby becoming a BIC.
\par
More recently, the existence of BIC's have been proved for quantum
dots coupled to reservoirs. Zero-width states have been found in
serial structures of dots or loops within simple, one-dimensional 
models of mesoscopic systems \cite{sdj94}. That BIC's are a general
phenomenon of quantum dots has been shown by Sadreev {\em et al.}, 
by modeling a quantum dot as a single billiard of variable shape 
attached to an external lead \cite{sbr06}. By resorting to an effective 
Hamiltonian approach to electron scattering through the device \cite{d00,op03},
the resonance features of the transmission probability have been related to 
the spectral properties of the closed quantum billiard. In particular, it has 
been found that zero-width states appear in the continuum of the open system, 
near the points of degeneracy or quasi-degeneracy of the closed-system 
eigenenergies. Two-dot systems have been studied as well, either
through analytically soluble  models allowing for a small number 
of discrete states in each dot \cite{sbr05}, or through direct numerical
solution of the two-dimensional Schr\"odinger equation via the contour
integration method \cite{onk06}. In the former case, BIC's appear
again when the eigenenergies of the closed system cross the transmission
zeros, with the electrons being trapped within the device \cite{sbr05}. 
In ref. \cite{onk06}, zero-width resonances have been analyzed in terms of 
the motion of the corresponding $S$-matrix poles in the complex energy plane. 
In analogy to what has been found for the usual coupled--channel problem
\cite{fw85i,fw85ii}, the onset of zero--width states is associated
to couples of poles, moving counterclockwise in the energy plane as the
bridge length increases, one of the poles touching the real energy axis
at critical values of this parameter. Moreover, BIC's appear at nearly 
periodic distances between the dots, when the length of the connecting 
bridge is varied.
\par
The aim of the present paper is to extend the analysis of resonances,
and of their evolution into BIC's as some continuous parameter of the 
system varies, to {\em two-dimensional} serial structures of dots. As is 
well--known, when several identical elements, such as dots, rings or 
constrictions, are connected in series, a band structure emerges in the 
transmission coefficient. In the same way as a band structure for 
electrons appears in solid--state physics \cite{lo99,pc02}, one has 
alternating regions of allowed and of essentially zero transmission as a 
function of energy. A noteworthy feature of periodic mesoscopic systems is 
that this ``miniband'' structure appears even for a relatively small number 
$(3\div 5)$ of components \cite{lo99}. We numerically solve the 
two--dimensional Schr\"odinger equation by a combination of mode--matching 
and $S$-matrix techniques, the total scattering operator for the device being 
obtained from the $S$--matrices referring to the various segments of the 
system through the $\star$--product composition rule \cite{da95,fg97,noi07}.
This approach provides numerically stable results for both physical and
complex values of the energy, even when some dimension of the system is
large \cite{noi07}, and has been already employed by us to investigate Fano 
resonances and threshold phenomena in ballistic transmission through dots 
with impurities \cite{noi07,noi07s}. In studying zero--width states in serial
systems, three issues will be of particular concern to us; $\left. i\right)$
how robust these trapped states are when the number of segments increases;
in other words, given the occurrence of a BIC in, say, a two--dot system
for a given configuration, there is a similar state for the analogous 
several-dots device?; $\left. ii\right)$ whether and to what extent the close 
relationship among zero--width states and the eigenstates of the {\em closed} 
system found for a single dot (or a coupled--dot pair) extends to serial 
structures; $\left. iii\right)$ what happens when the translational symmetry 
of the system is broken in correspondence to a BIC configuration by varying the
dimensions of the central dot.
\par
The paper is organized as follows. In Section \ref{results} we recall the
main characteristics of our approach, and present the results for
serial systems of dots coupled to a common waveguide. The non--trivial
changes of the transmission coefficients as some parameter of the device
is varied will be given a simple, transparent interpretation in the
light of the motion of the resonance poles in the complex energy plane.
Our main conclusions will be summarized in Section~\ref{conc}.

\section{Zero-width states in multi-mode serial structures}
\label{results}
\par
\begin{figure}[ht]
\centerline{\includegraphics[width=12 truecm,angle=0]{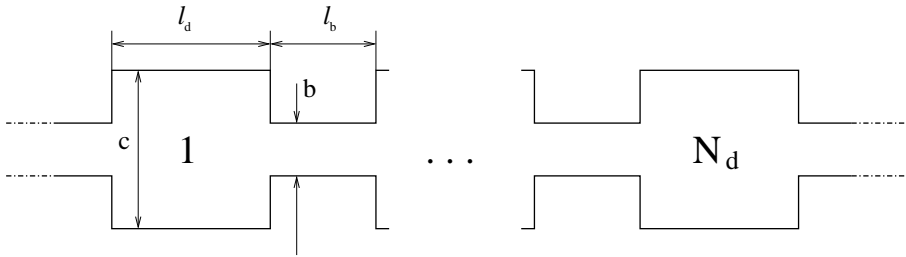}}
\caption{A serial structure of $N_d$ quantum dots coupled to an infinite
external lead.}
\label{fig1}
\end{figure}

We shall consider the device illustrated in Figure \ref{fig1}. The
quantum dots are modeled as rectangular cavities of total width
$c$ and length $l_d$, connected through bridges of width $b$ and
length $l_b$. The whole system is coupled to a uniform guide of
indefinite length, having the same width as the connecting necks.
In the ballistic regime, the electronic transport through the device
can be described as a scattering process, and the conductivity of the
quantum circuit can be expressed in terms of the transmission coefficients
of the system \cite{da95,fg97}. To evaluate these quantities, we start from 
the two--dimensional Schr\"odinger equation
\begin{equation}
\left \{ - \frac{\hbar^2}{2 m^{\ast}}\nabla^2_{\scriptscriptstyle{2}}\right \}
\Psi (x, y) = E \Psi (x, y)~~,
\label{scho1}
\end{equation}
where $\nabla^2_{\scriptscriptstyle{2}}$ represents the two--dimensional
Laplace operator, $E$ is the total energy, and $m^{\ast}$ is the electron's
effective mass in the conduction band.    
Eq. \ref{scho1} has been solved by a suitable combination of mode--matching
and $S$--matrix techniques, as detailed in Refs. \cite{noi07,noi07s}, to
which we refer the reader for details. Expanding the total wave--function
into complete sets of transverse--mode eigenfunctions in the various
segments (dots and bridges) of the device, Eq.~\ref{scho1} is reduced
to a set of one--dimensional Schr\"odinger equations for the expansion
coefficients, which depend upon the propagation variable $x$ only. These
coefficients are written in terms of forward and backward propagating
waves, with amplitudes which can be related to one another matching the
wave function and its first derivative at the various interfaces delimiting
the necks from the dots and the whole device from the external ducts. The
scattering operator for each segment is given once the amplitudes of 
the waves leaving an interface are expressed linearly in terms of the
coefficients of the incoming waves. The total $S$--matrix 
\begin{equation}
\mathbf{S} = 
\left(
\begin{matrix}
\mathbf{S}_{11}~ & ~\mathbf{S}_{12} \\
\noalign{\vspace{5 pt}}
\mathbf{S}_{21}~ & ~\mathbf{S}_{22}
\end{matrix}
\right) 
\label{smat}
\end{equation}
is finally obtained from the partial scattering operators through a recursive 
application of the $\star$--product composition rule \cite{da95,noi07}. In so 
doing, a different number of modes can be introduced in the various segments
of the device, to improve convergence, and evanescent modes can be taken into 
account while preserving the numerical accuracy of the calculation, even when
some dimension of the system gets large. The transmission coefficients are
contained in the matrix blocks $\mathbf{S}_{12}$ and $\mathbf{S}_{21}$;
more precisely, $\left(\mathbf{S}_{21}\right)_{nm}$ represents the
transmission coefficient to mode $n$ on the right of the device for an
electron impinging from the left in mode $m$, whereas 
$\left(\mathbf{S}_{12}\right)_{nm}$ is the transmission coefficient to the
final mode $n$ on the left from the initial mode $m$ on the right. The
sub-matrices $\mathbf{S}_{11}$ and $\mathbf{S}_{22}$ contain, on the other
hand, the corresponding reflection coefficients to the left and to the
right. 
\par
The scattering problem has been solved for an electron impinging from the 
left, with outgoing reflected and transmitted waves in all the open channels, 
and the total conductance $G$ (in units $2e^2/h$) has been evaluated through 
the two--probe B\"uttiker formula \cite{da95,fg97}
\begin{equation}
G = \sum_{m,n}\frac{k^{(l)}_n}{k^{(l)}_m}
\left|\left(\mathbf{S}_{21}\right)_{nm}\right|^2~~,
\label{condu}
\end{equation}
where the sum is restricted to the open channels only, and the propagation
wave numbers $k^{(l)}_n$ in the external leads are related to the total 
energy $E$ by \cite{noi07}
$$
k^{(l)}_n\equiv \
\sqrt{\frac{2m^\ast}{\hbar^2} E - \left(\frac{n\pi}{b}\right)^2}~~.
$$ 
As in our previous papers \cite{noi07,noi07s}, convergence is fully achieved 
when four channels are included in the leads, and up to 10 channels are taken
into account in the dots; fairly good results are already obtained, however,
with $2$ and $5$ modes in the lead and in the dots, respectively.
For the evanescent modes, the propagation wave-numbers are taken as purely 
imaginary quantities, with a positive coefficient, namely
$\displaystyle{k^{(l)}_n = i\sqrt{\left(n\pi/b\right)^2-2m^\ast E/\hbar^2}}$.
\par
\begin{figure}[h]
\centerline{\includegraphics[width=10 truecm,angle=-90]{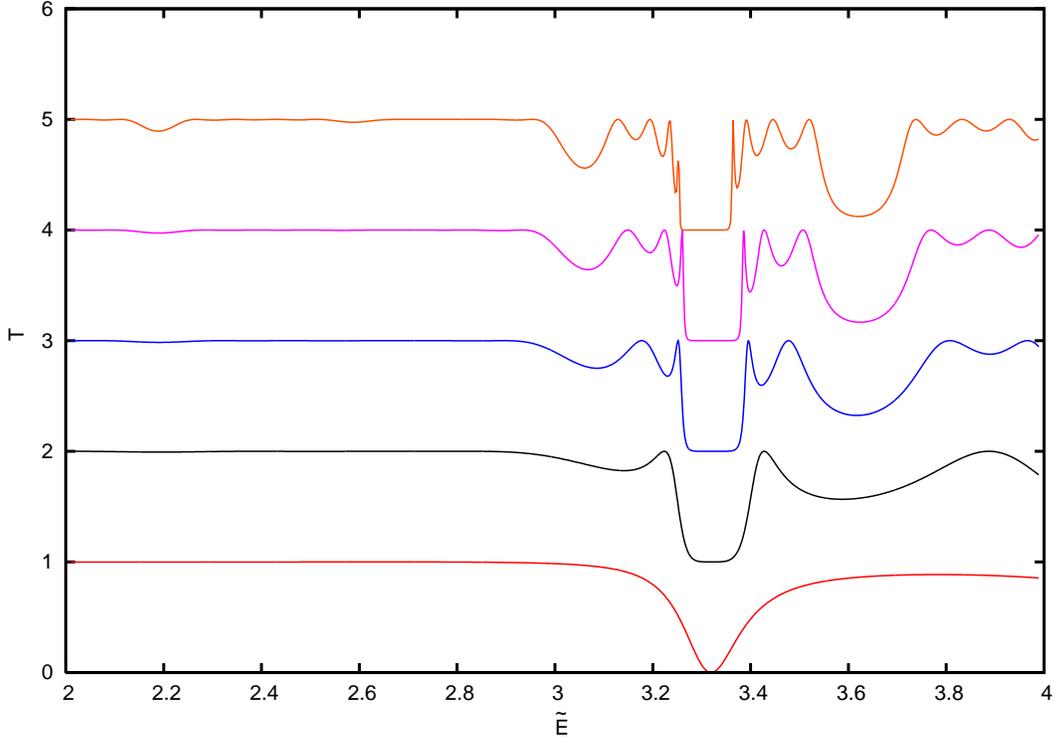}}
\caption{Total transmission as a function of the adimensional energy for 
increasing number of dots. From bottom to top $N_d=1,2,3,4,5$. Consecutive 
curves are vertically offset for clarity.}
\label{fig2}
\end{figure}

In figure \ref{fig2} we plot the conductance as a function of the energy 
for various serial devices, with an increasing number $N_d$ of dots. For
the sake of simplicity, we have limited ourselves to the first subband, 
where only one propagating mode is active. To have the results 
independent from the actual size of the device, we measured all lengths
in terms of the waveguide width $b$, and energies with respect to the
waveguide fundamental mode $\epsilon^{(l)}_1 \equiv \frac{\hbar^2}{2m^\ast}
\left(\frac{\pi}{b}\right)^2$. Adimensional quantities will be denoted 
with the ``tilde'' symbol, so that the various thresholds $\epsilon^{(l)}_n =
n^2\epsilon^{(l)}_1$ are simply given by $\tilde{\epsilon}^{(l)}_n = n^2$,
with $n=1,2,3,\ldots$. The calculations reported in Figure \ref{fig2}
refer to devices with a bridge length $\tilde{l}_b \equiv l_b/b
= 3.9$. The dots are modeled by square cavities with $\tilde{c} = \tilde{l}_d 
= 2$, symmetrically coupled to the bridges as well as to the waveguide. 

\par
As $N_d$ increases from 2 to 5, one sees $1, 2, 3, 4$ peaks on each side
of the transmission zero at $\tilde{E} \sim 3.3$. One moreover observes a 
more and more structured conductance profile at the upper edge of the band, 
and a smooth plateau of maximum transmission in the low--energy part of the 
considered energy region. The most interesting feature of Figure \ref{fig2} 
is the multiplet of peaks around the transmission zero at $\tilde E \sim 3.3$, 
having a position almost independent upon the number of coupled dots. The
increase in the number of oscillations is due to the more and more complex 
interference pattern among the waves reflected and transmitted at the dots, 
and has been previously found both in one--dimensional serial structures 
\cite{sdj94,ls93,zzyy99}, and in periodic, two-dimensional waveguides with 
stubs and constrictions \cite{ns98,cxy06,cxy07}.
\par
\begin{figure}[ht]
\centerline{
\includegraphics[width=10. truecm,angle=-90]{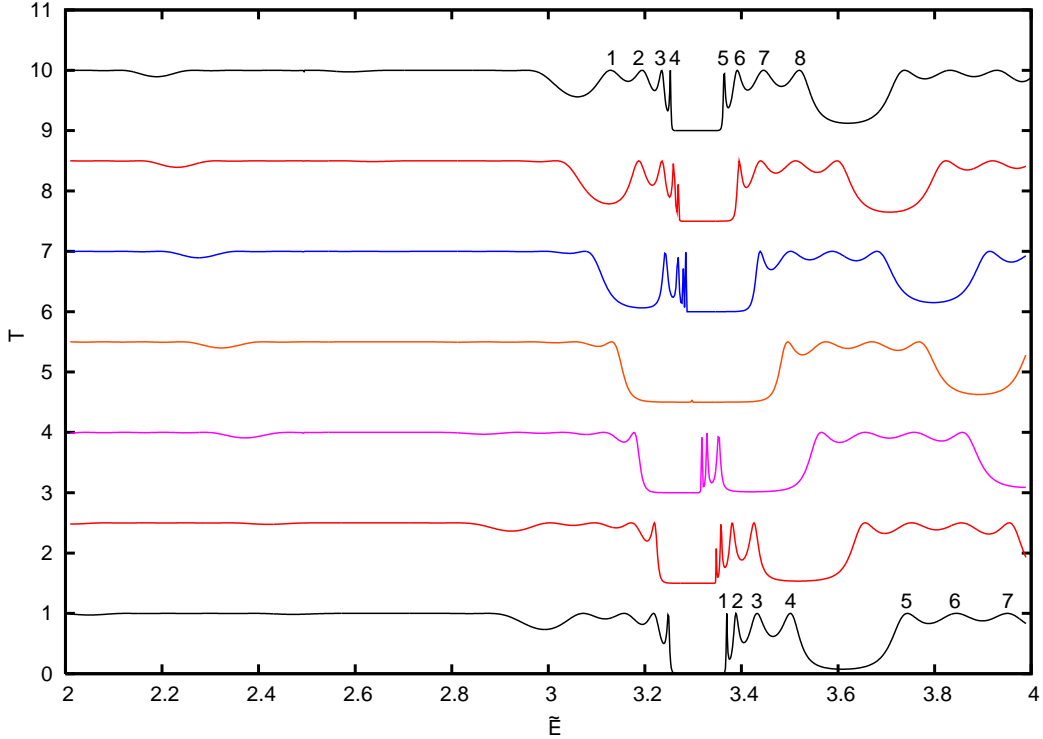}}
\caption{Total conductance of $5$ square dots symmetrically coupled to
a waveguide as a function of the adimensional energy. Each dot has
$\tilde{c} = \tilde{l}_d = 2.0$. From bottom to top the bridge length
$\tilde{l}_b$ increases from $\tilde{l}_b=3.3$ up to $\tilde{l}_b=3.9$
with $\Delta \tilde{l}_b=0.1$. The numbers $1,\ldots,8$ label the peaks
to which we refer in the text. Consecutive curves are vertically offset 
for clarity.}
\label{fig3}
\end{figure}

In figure \ref{fig3} we plot (from bottom to top) the conductance of
a 5-dot device in the first energy subband, for $\tilde{l}_b$ increasing
from $3.3$ up to $3.9$, in steps of $0.1$. As $\tilde{l}_b$ gets longer, 
the resonance peaks move towards lower energies, the typical binding effect 
one observes when some characteristic length of the system increases 
\cite{noi07}. At the same time, the four peaks on the right of the 
transmission zero (numbered from $1$ to $4$ in figure \ref{fig3}) 
approach each other and the transmission minimum, until they disappear for 
$\tilde{l}_b=3.6$, where a zero-width state is produced. 
For longer bridge lengths, they appear again on the left of the transmission 
minimum, with increasing relative distances. The four peaks on the left of 
the transmission zero for $\tilde{l}_b=3.3$, on the other hand, have in the 
mean time merged into the low energy background discernible in figure 
\ref{fig3}. One observes also peaks moving down in energy, coming 
from the second scattering threshold, at $\tilde E = 4$. For $\tilde{l}_b=3.9$
a conductance profile quite similar to the initial one can be observed. 
The same trend  is observed for a different number of dots, with the BIC 
occurring more or less at the same energy position and almost for the same 
value of the bridge length.  
\par
\begin{figure}[ht]
\centerline{
\includegraphics[width=10. truecm,angle=-90]{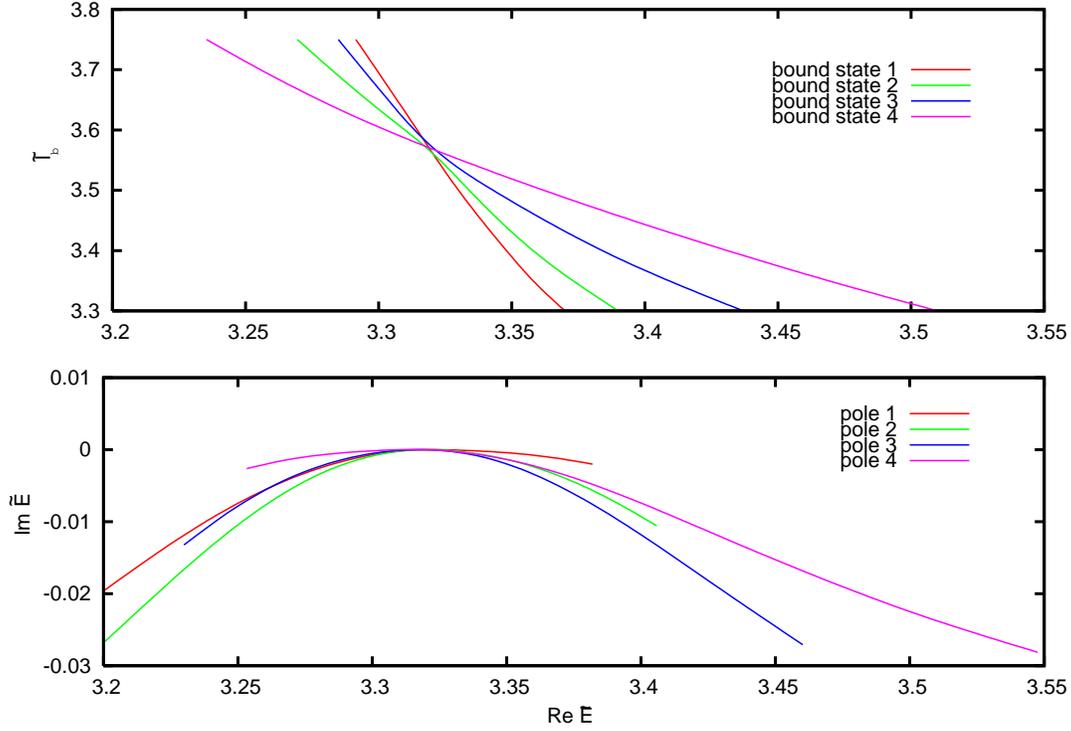}}
\caption{Lower panel: counterclockwise motion of the poles associated to the
peaks $1,2,3,4$ of Figure \ref{fig3} in the complex energy plane 
for $\tilde{l}_b$ increasing from $3.3$ up to $3.9$. Upper panel: 
the associated eigenenergies of the corresponding closed system in the
$(\tilde{E},\;\tilde{l}_b)$ plane.} 
\label{fig4}
\end{figure}
\begin{figure}
\centerline{
\includegraphics[width=10. truecm,angle=-90]{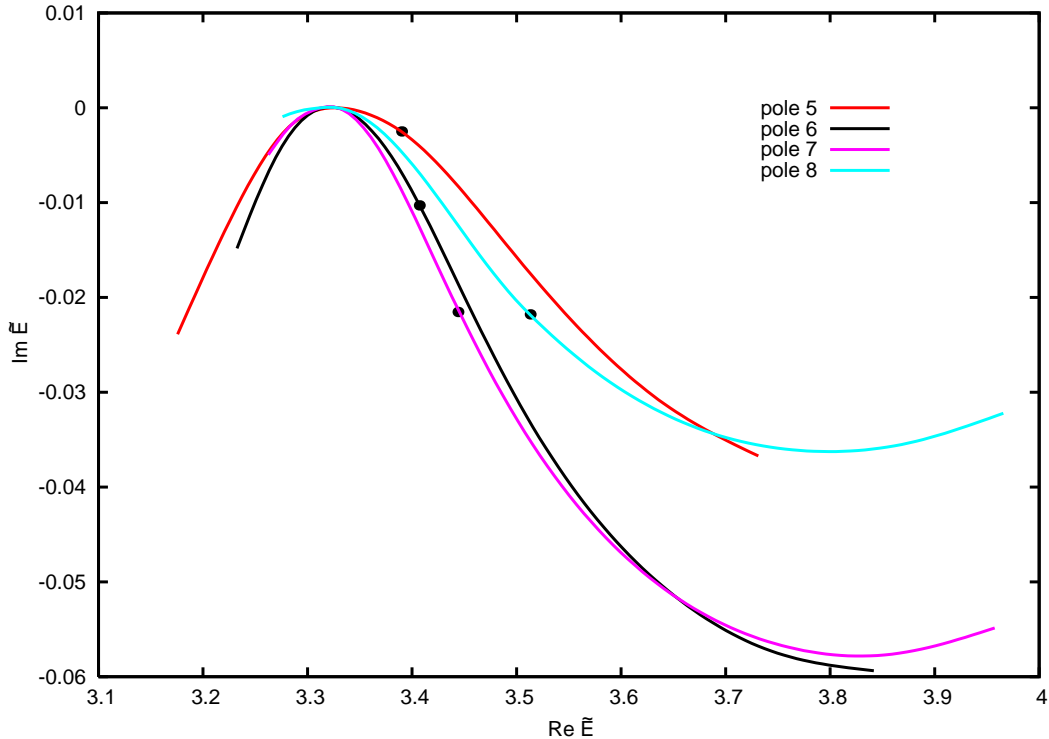}}
\caption{Counterclockwise motion of the poles $5,6,7,8$ of Figure
\ref{fig3} in the complex energy plane with $l_b$ increasing from
$\tilde{l}_b=3.3$ up to $\tilde{l}_b=4.5$. The full dots mark the
pole positions for $\tilde{l}_b=3.9$.}
\label{fig5}
\end{figure}

As Figure \ref{fig3} exhibits, the behavior of the conductance as a function 
of energy with varying bridge length is by far non trivial. A simpler picture
emerges, when one considers the motion of the $S$--matrix poles in the
multi--sheeted energy Riemann surface. For the present purposes, one
can limit oneself to the Riemann sheet where ${\rm Im} k^{(l)}_1 < 0$
whereas ${\rm Im} k^{(l)}_j > 0$ for $j > 1$; there, one has exponentially
diverging waves in the open channel, and exponentially decreasing waves
in the other channels. A pole at $E_p \equiv E^{(R)}-i\Gamma$ can be
associated to a resonance in the conductance $G$ around $E\sim E^{(R)}$,
of width $2\Gamma$, provided that $E^{(R)}$ is in between the first
and second scattering threshold \cite{noi07s,bkps82}. We applied our
approach to solve the Schr\"odinger equation for complex energies and
channel momenta, so as to locate the $S$--matrix poles in the relevant
sheet. The results are given in Figures \ref{fig4} and \ref{fig5}. In the 
lower panel of the former we give the four poles corresponding to peaks 
$1\div 4$ in figure \ref{fig3}, while in the latter we plot the poles 
associated to peaks $5\div 8$. As $\tilde{l}_b$ increases from 
$\tilde l_b = 3.3$ up to $\tilde{l}_b =3.6$, the poles of figure \ref{fig4} 
rotate counterclockwise in the fourth quadrant of the energy plane approaching 
the real energy axis, where they collapse and coincide with the transmission 
zero for $\tilde{l}_b=3.6$. As the bridge length increases further, the four 
poles move away from the energy axis, so that the corresponding peaks are again
discernible in the conductance at lower and lower energies. 
As for the poles at the upper edge of the transmission band, they move 
monotonically downwards in the energy plane towards the energy axis until 
they produce peaks $5\div 8$ at the right of the transmission minimum, 
which can be seen in figure \ref{fig3} for $\tilde{l}_b=3.9$. In the mean
time, new poles come into play through the second scattering threshold, 
coming from the higher energy region. As far as these poles have $E^{(R)}>4$,
however, they cannot produce observable effects on the conductance; indeed, 
transmission resonances above the second scattering threshold are due to 
poles residing on other, different sheets of the energy Riemann surface 
\cite{noi07s}. Stated in the language of dispersion theory, as the bridge 
length increases one observes that some of the $S$--matrix poles undergo a 
transition from a ``shadow" state to a ``dominant'' role \cite{noi07s,et64}. 
We verified also that the low energy plateau in the conductance 
corresponds to a group of closed packed poles in the energy plane. 
As $\tilde{l}_b$ increases, the four poles producing the peaks on the left of 
the transmission zero for $\tilde{l}_b=3.3$ join these low energy poles so 
that the corresponding peaks disappear in the background. 
\par
For greater bridge lengths a zero--width state appears again at 
$\tilde E\sim 3.3$ in the conductance spectrum. This is exhibited in the
left-most part of Figure \ref{fig5}, where we give also the pole positions 
for $\tilde{l}_b \geq 3.9$; for $\tilde{l}_b=4.22$ the four poles 
coincide with each other and with the transmission zero, and a new BIC 
appears. Note that the four poles associated to the BIC at $\tilde{l}_b=
3.3$ have in the meantime moved downward in energy towards the structureless
background.
It is finally worth to mention what happens in terms of the $S$--matrix
poles when the number of dots increases for a given bridge length. With
reference to Figure \ref{fig1}, the dip in conductance one observes 
for $N_d=1$ is due to a pole at $E_p \simeq 3.31-0.072i$; no other pole
is found near the energy axis in the considered energy region. Things
are already quite different when passing to the two--dot case; the smooth 
plateau of maximum transmission at low--energy is associated to four 
closed--packed poles with $2.03 < E^{(R)} < 3.02$ and 
$0.303 < \Gamma < 0.379$. As $N_d$ increases, more and more strongly coupled 
poles appear in the low-energy part of the first transmission band. At the 
same time, more and more poles appear very near the energy axis, on both sides 
of the transmission zero. In all instances, the transmission zero occurs at 
$\tilde{E} \simeq 3.32$.
\par 
For coupled--channel problems in Atomic Physics, the dynamics underlying
the occurrence of BIC's has been investigated by Friedrich and Wintgen  
in the framework of a three--channel model \cite{fw85i,fw85ii}. Assuming 
one open and two closed channels, where the wave function is taken to be 
proportional to the bound-state solutions of the corresponding uncoupled 
Schr\"odinger equations, one can solve for the scattering component so as to 
obtain explicit expressions for the resonance positions and widths. Friedrich 
and Wintgen were able to show that, as the bound state energies $E_1$ and $E_2$
were tuned varying some characteristic parameter, the interference effects 
lead to an avoided crossing of the resonance positions. At the same time, one 
observes dramatic changes in the resonance widths, one of them vanishing when 
$E_1=E_2$ and the bound states in the uncoupled closed channels become 
degenerate. A strict connection between the occurrence of zero-width 
resonances and the presence of degenerate eigenenergies for the corresponding
{\em two-dimensional closed} system has been also found for a quantum 
billiard of variable shape by Sadreev {\em et al.} \cite{sbr06}. The two-dot 
case has been considered in Ref. \cite{sbr05}. There too, the trapping of the 
particle in the internal bridge occurs in correspondence to the 
crossing of the transmission zero by the eigenenergies of the closed system.
We ascertained whether this simple relation between the occurrence of
zero--width states and the eigenenergies of the closed system still
holds in the serial structures under consideration. To this end, one
has to solve the Dirichlet boundary--value problem for periodic domains
such that of Figure \ref{fig1}. We accomplished this through a
modification of the $S$--matrix approach. The condition that the full
wave--function vanishes on the leftmost and rightmost edge of the closed 
system implies that the amplitudes of the forward and backward propagating 
waves have to be related by
\begin{equation}
\overrightarrow{c}^{(1)}_n = - \overleftarrow{c}^{(1)}_n \quad
\overrightarrow{c}^{(N_d)}_n = - e^{-2i\tilde{k}^{(d)}_n \tilde{l}_d}
                               \overleftarrow{c}^{(N_d)}_n~~,
\label{phase}
\end{equation}
where $\overrightarrow{c}^{(1)}_n$ and $\overleftarrow{c}^{(1)}_n$ are
the amplitudes for the waves propagating in the first dot to the right 
and to the left, respectively, and $\overrightarrow{c}^{(N_d)}_n$,
$\overleftarrow{c}^{(N_d)}_n$ the corresponding quantities in the 
last dot on the right. By $\tilde{k}^{(d)}_n$ we denote the adimensional 
propagation wave numbers in the various channels inside the dots 
\cite{noi07}. Note that, even if these relations have been written for
an open channel, they apply in the closed channels also, so that convergence
with respect to the expansion into transverse basis functions is
guaranteed when solving the Dirichlet problem in the domain. Now, by
definition of the $S$--matrix, one can write in obvious matrix notation 
\begin{equation}
\left(\begin{array}{c}
\overleftarrow{\mathbf c}^{(1)} \cr
\noalign{\vspace{5pt}}
\overrightarrow{\mathbf c}^{(N_d)}
\end{array}\right)=
\left(\begin{array}{cc}
{\mathbf S}_{11} & {\mathbf S}_{12} \cr
\noalign{\vspace{5pt}}
{\mathbf S}_{21} & {\mathbf S}_{22}
\end{array}\right)
\left(\begin{array}{c}
\overrightarrow{\mathbf c}^{(1)} \cr
\noalign{\vspace{5pt}}
\overleftarrow{\mathbf c}^{(N_d)}
\end{array}\right)~~.
\label{srel}
\end{equation}
Combining Eqs. \ref{phase} and \ref{srel} one obtains
\begin{equation}
{\mathbf {\cal B}}\overrightarrow{\mathbf c}^{(N_d)} = 0~~,
\label{eigen}
\end{equation}
where 
\begin{equation}
\mathbf{\cal B} \equiv {\mathbf S}_{21}\left({\mathbf 1}+
{\mathbf S}_{11}\right)^{-1}{\mathbf S}_{12}{\mathbf \Omega}
-{\mathbf S}_{22}{\mathbf \Omega} - {\mathbf 1}~~,
\label{bi}
\end{equation}
with ${\mathbf \Omega}_{mn} \equiv e^{2i\tilde{k}^{(d)}_m\tilde{l}_d}
\delta_{mn}$. Eq. \ref{eigen} represents a system of homogeneous equations for 
the amplitudes $\overrightarrow{c}^{(N_d)}_n$, which has a non--trivial 
solution if and only if the condition ${\rm det}{\mathbf {\cal B}}=0$ is 
satisfied. This condition can be regarded as the secular equation fixing the 
eigenenergies of the standing waves in the closed, periodic domain under
consideration. Since the energy enters in a highly non--trivial way in the
secular equation through the channel wave-numbers, we actually evaluated
${\rm det}{\mathbf {\cal B}}$ for varying energies, and looked for the
zeros of its modulus through a minimization procedure. We tested our 
approach for a rectangular domain, whose eigenenergies are known analytically, 
and for the quarter square Sinai billiard, which has been studied by 
different means in Ref. \cite{zvl08}. We found excellent agreement in both 
cases when the same number of basis functions as in the scattering calculations
were included.
\par 
In the upper panel of Figure \ref{fig4} we exhibit the results of our 
calculations for the $5$-stub case. The eigenenergies are plotted in
the $(\tilde{E},\tilde{l}_b)$ plane, and refer to the states associated with 
peaks $1\div 4$, occurring when the system is coupled to the waveguide. From 
a comparison with the motion of the corresponding $S$-matrix poles 
(lower panel) one sees that, as the resonance poles move towards the energy 
axis and overlap the transmission zero in correspondence to the BIC, the  
eigenenergies of the closed system become closer and closer to each other, 
until one has four degenerate eigenvalues for $\tilde{l}_b=3.6$. We verified 
that for three or four coupled dots one has a pair or a triplet of degenerate 
eigenenergies of the closed structure in correspondence to the BIC. The 
zero-width configuration occurs moreover practically at the same value of 
$\tilde{l}_b$. Overall, our findings show that the occurrence of trapped 
states with zero width is a {\em robust effect} in serial structures. 
\par
\begin{figure}[ht]
\centerline{
\includegraphics[width=10. truecm,angle=-90]{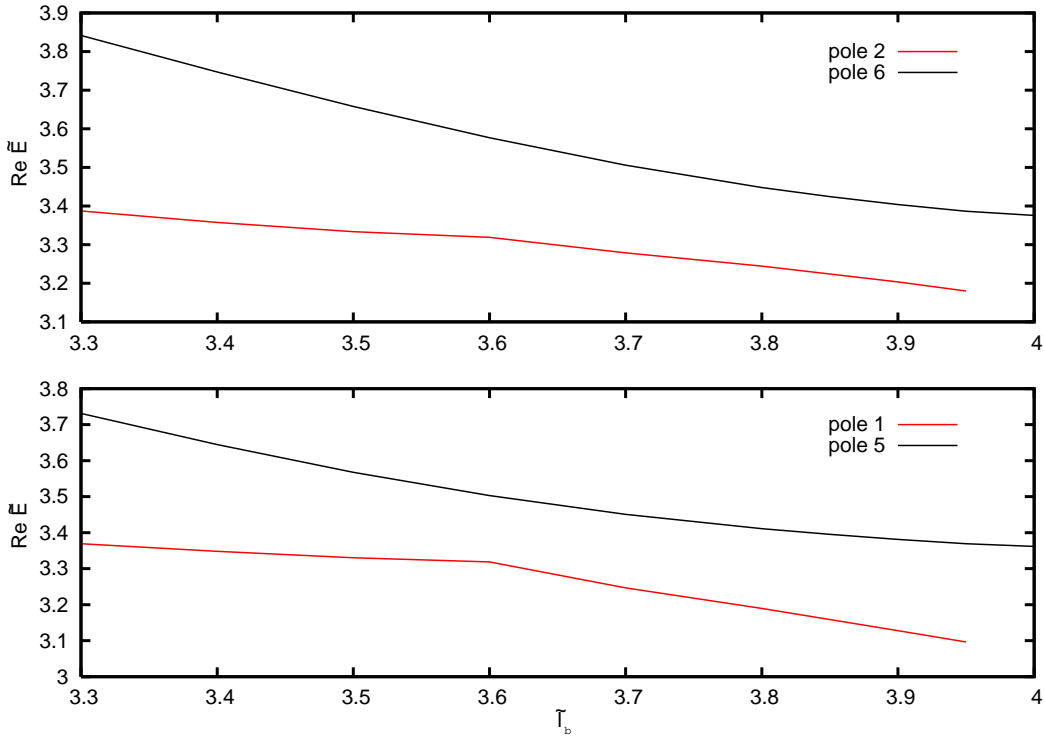}}
\caption{Position of the resonances $1$ and $5$ as a function of 
$\tilde{l}_b$ for the $5$-dot system of Figure \ref{fig3} near the BIC 
configuration (lower panel). The same graph is given for resonances 
$2$ and $6$ in the upper panel.}
\label{fig6}
\end{figure}

We looked also for possible avoided crossings between the trajectories of
the resonance poles. In Figure \ref{fig6} we give the real part of the pole
energies $\tilde{E}_p$ for the pairs of poles $(1,\;5)$ and $(2,\;6)$ 
in the $(\tilde{E},\tilde{l}_b)$ plane. The avoided crossing is clearly 
discernible for the lowest-energy resonances, as can be inferred from the lower
panel of Figure \ref{fig6}; for $\tilde{l}_b=3.6$, where the width of 
one of the resonances is zero, the relative distance among the corresponding 
poles reaches a minimum, and then increases again. This behavior is less
evident for the other resonance, as the upper panel of Figure \ref{fig6}
shows, and deteriorates further as one considers the resonance poles at higher 
energies. This effect can be attributed to the presence of other, nearby poles
at the upper edge of the transmission band. In particular, as the bridge length
increases, more poles come into play from higher energies through the
second scattering threshold, and may perturb the motion of the high--lying
resonance poles of the upper multiplet. 
\par
We finally investigated the effect of a symmetry breaking of the
system on the BIC. To be definite, we refer again to the $5$--dot 
case with $\tilde{l}_b=3.6$. When the central dot is made different
from the lateral ones, the quadruplet of poles at the transmission zero 
are no longer degenerate, nor they coincide with the transmission zero. This 
removal of the BIC when the symmetry among the various dots is slightly 
broken is consistent with the results of ref. \cite{sbr05} for the two-dot 
case. The general rule according to which there is an antibinding effect, 
and the poles move in a counterclockwise way as some characteristic 
length of the system increases, is still obeyed. The four poles, however, 
move in the complex energy plane with different velocities, producing a 
multiplet of narrow peaks which depend in a rather complex way upon the 
bridge length. In particular, when the central dot is made shorter with
respect to the external ones, two poles move faster towards higher energies 
leaving the BIC position, until they give rise to peaks strongly coupled to 
the resonances residing at the upper edge of the conduction band. The other 
two poles of the quadruplet, on the contrary, do not leave the region of the 
transmission zero, and give rise to very narrow peaks in the limiting 
situation of a central dot of the same height as the connecting bridge,
so that one has two couples of dots connected by a central bridge with
$\tilde{l}_b = 9.2$.  

\section{Conclusions}
\label{conc}
\par
In this paper we have considered the occurrence of zero--width states
in the continuum of periodic systems of several coupled dots opened into
an external waveguide. The present analysis can be therefore considered
as the extension to serial devices of what has been done for two--dot
systems \cite{sbr05,onk06}. We studied how the transmission properties
change as the length $\tilde{l}_b$ of the connecting bridges varies,
and found that the BIC phenomenon is a rather robust effect with respect
to the number of dots $N_d$. When a BIC is produced for a suitable value 
of $\tilde{l}_b$ in a two--dot system, it persists for a larger number 
$N_d$ of dots. Even if we 
limited ourselves to present detailed results for a $5$--dot device, 
we verified that our conclusions still hold up to ten coupled dots.
Overall, with varying $\tilde{l}_b$ the conductance profile varies in
a non--trivial way. A much simpler picture emerges, however, when
one looks at the trajectories of the resonance poles on the
relevant sheet of the complex energy plane. As $\tilde{l}_b$ increases,
the poles move counterclockwise in the energy plane; when a multiplet
of poles collapses onto a transmission zero on the real energy axis
the electron is trapped inside the device and a zero--width state
emerges. This peculiar configuration is removed once the translational 
symmetry of the system is broken. Finally, we have shown that the
present $S$--matrix approach can be modified so as to treat the
Dirichlet boundary--value problem for the closed system. In close
analogy with what has been found for a single \cite{sbr06}, or two 
coupled dots \cite{sbr05}, we found that in correspondence to a BIC
a whole multiplet of eigenenergies of the closed system is
degenerate with the transmission minimum.

\end{document}